\documentclass[aps,pra,amsfonts,amsmath,showpacs]{revtex4}

\begin{document}

\title{The Optimal Single Copy Measurement for the Hidden Subgroup Problem}
\author{Dave Bacon}
\affiliation{Department of Computer Science \& Engineering, University of Washington, Seattle, WA 98195}
\author{Thomas Decker}
\affiliation{Department of Computer Science \& Engineering, University of Washington, Seattle, WA 98195}
\date{July 30, 2007}
\begin{abstract}
The optimization of measurements for the state distinction problem has recently been applied to the theory of quantum algorithms with considerable successes, including efficient new quantum algorithms for the non-abelian hidden subgroup problem.  Previous work has identified the optimal single copy measurement for the hidden subgroup problem over abelian groups as well as for the non-abelian problem in the setting where the subgroups are restricted to be all conjugate to each other.  Here we describe the optimal single copy measurement for the hidden subgroup problem when all of the subgroups of the group are given with equal a priori probability.  The optimal measurement is seen to be a hybrid of the two previously discovered single copy optimal measurements for the hidden subgroup problem.
\end{abstract}
\maketitle

\section{Introduction}

The hidden subgroup problem (HSP, defined below) is one of the most widely studied problems in the theory of quantum algorithms.  The reasons for this are twofold.  On the one hand, Shor's efficient quantum algorithms for factoring and for computing the discrete logarithm can both be seen as instances of the abelian HSP~\cite{Shor:94a}.  On the other hand, efficient algorithms for non-abelian HSPs would allow for efficient algorithms for the graph isomorphism problem~\cite{Beals:97a,Boneh:95a,Hoyer:97a,Ettinger:99a} and certain unique shortest vector in a lattice problems~\cite{Regev:04a}, two problems that currently do not admit efficient classical or quantum algorithms.  This latter fact has motivated a great deal of research into whether quantum computers can efficiently solve the non-abelian HSP, with notable successes~\cite{Hallgren:00a,Grigni:00a,Ivanyos:01a,Gavinsky:04a,Friedl:03a,Kuperberg:03a,Moore:04a,Bacon:05a}, but with no success in finding efficient quantum algorithms for the HSP over groups that would yield efficient algorithms for the graph isomorphism and the shortest vector problems.

Recently an approach to solving the non-abelian HSP has emerged by recasting the problem as a state distinction problem~\cite{Ip:03a,Bacon:06c,Bacon:05a}.  In this setting, optimal measurements for solving the problem can be derived and in some special cases these measurements can be efficiently implemented and lead to an efficient quantum algorithm for the relevant HSP.  Thus, for instance, this approach has led to an efficient quantum algorithm for the HSP over the Heisenberg group and a natural generalization of this group~\cite{Bacon:05a}.  The first to consider optimal measurements for the HSP was Ip ~\cite{Ip:03a}, who considered this problem in the abelian setting when all subgroups are given with equal a priori probability.  Ip derived an optimal measurement for the HSP in this case and showed that it corresponds to Shor's algorithm.  Explicit optimal measurements for HSPs over the dihedral group~\cite{Bacon:06c} and the Heisenberg group~\cite{Bacon:05a} were then derived by Bacon, Childs and van Dam.  It was observed that in both of these cases the optimal measurement was the so-called pretty good measurement (PGM)~\cite{Hausladen:94a}.  Building on this result, Moore and Russell~\cite{Moore:05c} showed that the PGM is the optimal single copy measurement for the non-abelian HSP when all of the subgroups are conjugate to each other and is the optimal multi-copy measurement for the non-abelian HSP when the subgroups are all Gel'fand pairs with the group~\cite{Moore:05c}.  Here we derive the optimal single copy measurement for the HSP when all of the subgroups are given with equal probability, removing the restriction of Moore and Russell that these subgroups are conjugate to each other.  The measurement we derive can be seen as a hybrid between the measurement of Ip~\cite{Ip:03a} and the PGM~\cite{Bacon:06c,Moore:05c}.

\section{The Hidden Subgroup Problem and Optimal Measurements}

We begin by defining the hidden subgroup problem.  A detailed review of this problem can be found in ~\cite{Lomont:04a}.
\begin{quote}
    {\bf Hidden Subgroup Problem (HSP)} Let ${\mathcal G}$ be a group and $f$ a function from elements of this group to some set $S$.  This function is promised to be constant and distinct on different left cosets of a subgroup ${\mathcal H}$ of ${\mathcal G}$.  The hidden subgroup problem is to identify, by querying $f$, the subgroup ${\mathcal H}$.
\end{quote}
In all of the cases we work with, ${\mathcal G}$ will be finite, and elements of ${\mathcal G}$ and of $S$ will be expressed as $O({\rm poly} \log|{\mathcal G}|)$ bitstrings.  An algorithm for the HSP is considered to be efficient when the running time is polynomial in $\log|{\mathcal G}|$.  By identifying ${\mathcal H}$ we do not require that all of the elements of ${\mathcal H}$ be returned, but only that a generating set or any succinct description of the subgroup be returned.  When ${\mathcal G}$ is abelian, we call this problem the abelian HSP and similarly when ${\mathcal G}$ is non-abelian we call this problem the non-abelian HSP.

In the standard query model of the hidden subgroup problem, one approaches a quantum algorithm for the HSP by querying the function $f$ in superposition over the group elements.  This produces the state
\begin{equation}
{1 \over \sqrt{|{\mathcal G}|} } \sum_{g \in {\mathcal G}} |g\rangle \otimes |f(g)\rangle.
\end{equation}
The second register of this state (whose range is the set $S$) is then discarded and we obtain a mixed state depending only on the subgroup ${\mathcal H}$ hidden by $f$,
\begin{equation}
\rho_{\mathcal H}={|{\mathcal H}| \over |{\mathcal G}|} \sum_{g\in \widehat{\mathcal G}} |g{\mathcal H} \rangle \langle g{\mathcal H}|,
\label{eq:hsstate}
\end{equation}
where we have defined the left coset states
\begin{equation}
|g{\mathcal H}\rangle := {1 \over \sqrt{|{\mathcal H}|}} \sum_{h \in {\mathcal
H}} |gh\rangle,
\end{equation}
and $\widehat{\mathcal G}$ is a set of coset representatives for the subgroup ${\mathcal H}$.  We call $\rho_{\mathcal H}$ the hidden subgroup state.  In the standard approach to solving the hidden subgroup problem on a quantum computer, one is given access to one or many copies of $\rho_{\mathcal H}$.  The goal is then to perform a set of quantum gates and a measurement that efficiently identifies ${\mathcal H}$.

In the optimal measurement approach to the HSP, one examines the HSP from a state distinction perspective.  Suppose that the function $f$ hides ${\mathcal H}$ with probability $p_{\mathcal H}$.  Then in the single copy case, we can assume that we have been given $\rho_{\mathcal H}$ with probability $p_{\mathcal H}$.  One then attempts to find a generalized measurement that maximizes the probability of successfully identifying the hidden subgroup ${\mathcal H}$.  A generalized measurement can be described by a set of positive operators that sum to identity.  If $E_{\mathcal H}$ is the measurement operator corresponding to our guessing that the hidden subgroup is ${\mathcal H}$, then the optimal measurement problem in the single copy case maximizes the probability of successfully identifying the subgroups,
\begin{equation}
p_{\rm succ}=\sum_{{\mathcal H} \in {\rm Sub}({\mathcal G})} p_{\mathcal H} {\rm Tr}[\rho_{\mathcal H} E_{\mathcal H}],
\end{equation}
subject to $E_{\mathcal H}$ being a valid generalized measurement: $E_{\mathcal H} \geq 0$ and $\sum_{{\mathcal H} \in {\rm Sub}({\mathcal G})} E_{\mathcal H}=I$.  Here ${\rm Sub}({\mathcal G})$ is the set of subgroups of ${\mathcal G}$.   This is the well-known problem of state distinction.  The state distinction problem was considered in the seventies by Holevo ~\cite{Holevo:73b}, and Yuen, Kennedy, and Lax~\cite{Yuen:75a} where necessary and sufficient conditions for the optimal measurement were derived for the generic state distinction problem. Applied to the above formulation of the hidden subgroup problem, these necessary and sufficient conditions yield that a measurement $E_{\mathcal H}$ is optimal if and only if
\begin{eqnarray}
\sum_{{\mathcal H} \in {\rm Sub}({\mathcal G})} p_{\mathcal H} E_{\mathcal H} \rho_{\mathcal H}&=&\sum_{{\mathcal H} \in {\rm Sub}({\mathcal G})} p_{\mathcal H}  \rho_{\mathcal H} E_{\mathcal H} , \nonumber \\
\sum_{{\mathcal H} \in {\rm Sub}({\mathcal G})} p_{\mathcal H} E_{\mathcal H}\rho_{\mathcal H} &\geq& p_{{\mathcal H}^\prime} \rho_{{\mathcal H}^\prime}, \quad {\rm for~all~}{\mathcal H}^\prime \in {\rm Sub}({\mathcal G})\label{eq:optimal}.
\end{eqnarray}
We may also consider the multi-copy optimal measurement, where we query the function $r$ times.  In that case we produce $r$ copies of the hidden subgroup state $\rho_{\mathcal H}$, and the optimality condition is obtained by simply replacing $\rho_{\mathcal H}$ by $\rho_{\mathcal H}^{\otimes r}$ in Eq.~(\ref{eq:optimal}).

In ~\cite{Ip:03a}, Ip considered the question of identifying the optimal measurement for the abelian HSP when each of the subgroups of ${\mathcal G}$ is given with equal a priori probability.  Here we briefly review this result.  Define
\begin{equation}
P_{\mathcal H}:={|{\mathcal G}| \over |{\mathcal H}|} \rho_{\mathcal H}.
\end{equation}
The optimal single copy measurement for the abelian HSP is then given by the measurement operators (recursively defined)
\begin{equation}
E_{\mathcal H}: = P_{\mathcal H}-\sum_{{\mathcal J} \supset {\mathcal H}} E_{\mathcal J}.
\end{equation}
In the case of $r$ copies, the optimal measurement is given by the similar expression
\begin{equation}
E_{\mathcal H}^{(r)}:= P_{\mathcal H}^{\otimes r}-\sum_{{\mathcal J} \supset {\mathcal H}} E_{\mathcal J}^{(r)}.
\end{equation}
Finally Ip also showed that for the HSP over the dihedral group, the above expression is not the optimal measurement.

Following Ip's work, Bacon, Childs, and van Dam considered the optimal measurement for the HSP over the dihedral~\cite{Bacon:06c} and the Heisenberg groups~\cite{Bacon:05a}.  In both of these cases, the measurement turned out to be the pretty good measurement (PGM), so named because it does a good job of identifying states~\cite{Hausladen:94a}.  If the state $\rho_i$ is given with probability $p_i$, then the PGM is a measurement consisting of measurement operators $E_i$ defined as
\begin{equation}
E_i:=p_i M^{-1/2} \rho_i M^{-1/2}, \quad M:=\sum_{i} p_i \rho_i,
\end{equation}
where the inverse square root $M^{-1/2}$ is taken over the support of $M$. Bacon, Childs, and van Dam showed that the PGM was optimal for the dihedral and Heisenberg HSPs for both the single and multi-copy case and when the subgroups were restricted to certain order $2$ or order $p$ subgroups (which is enough to be able to solve the HSP over all subgroups due to a generalization of a theorem of Ettinger and H{\o}yer~\cite{Ettinger:00a,Bacon:05a,Bacon:06c}.)

Finally, Moore and Russell showed that under certain further conditions the PGM is guaranteed to be optimal~\cite{Moore:05c}.  In particular they showed that if the hidden subgroup is restricted to come from the set of subgroups conjugate to a fixed subgroup, then for the single copy case, if these subgroups are sampled with uniform probability the PGM is optimal.  Further Moore and Russell showed that the PGM is optimal for the multi-copy case when all of the hidden subgroups form Gel'fand pairs with the parent group.

It is easy to see that for the abelian HSP, the optimal measurement is not always the PGM.  For instance, suppose that the group is ${\mathcal G}={\mathbb Z}_2$ and the hidden subgroup is either the entire group, $\{e,r\}$ or the trivial subgroup, $\{e\}$ with equal probability.  Then the hidden subgroup states are simply
\begin{equation}
\rho_0={1 \over 2}(|e\rangle+|r\rangle)(\langle e|+ \langle r|), \quad \rho_1={1 \over 2} (|e\rangle \langle e|+|r\rangle \langle r|)
\end{equation}
Defining $|\pm\rangle:={1 \over \sqrt{2}}(|e\rangle\pm |r\rangle)$, we can express these as $\rho_0=|+\rangle \langle +|$ and $\rho_1={1 \over 2}(|+\rangle \langle +| +|-\rangle \langle -|)$.
It is then easy to calculate that the PGM has measurement operators
\begin{equation}
E_0={2 \over 3} |+\rangle \langle +|
\quad {\rm and} \quad E_1={1 \over 3} |+\rangle \langle +| + |-\rangle \langle -|.
\end{equation}
The probability of successfully identifying the subgroup is then $P_{\rm succ}={1 \over 2} {\rm Tr} [E_0 \rho_0+E_1 \rho_1]={2 \over 3}$.  However, if we had chosen the optimal measurements $F_0=|+\rangle \langle +|$ and $F_1=|-\rangle \langle -|$, then this succeeds with probability ${3 \over 4}$.  We therefore see that the PGM is not optimal.  We can thus conclude that while the PGM is optimal for a plethora of HSPs, it is not always so.  Therefore, an important unsolved question in the optimal measurement approach to the HSP is to determine the optimal measurement for the HSP beyond the cases where the PGM is optimal or Ip's measurement is optimal (see~\cite{Harrow:06a} and~\cite{Barnum:02a} for the near optimality of the PGM).

In this paper we derive the optimal single copy measurement for the HSP when the subgroups are chosen from all possible subgroups with equal probability.  We will show that this measurement is a hybrid between the PGM and the measurement procedure elucidated by Ip.

\section{Optimal Single Copy Hidden Subgroup Problem Measurement}

We begin by reviewing properties of hidden subgroup states and then turn to identifying the optimal measurement.  We find the optimal measurement by choosing a measurement ansatz that is a hybrid between the PGM and the measurement described by Ip ~\cite{Ip:03a}.

\subsection{Properties of the Hidden Subgroup State}

The hidden subgroup state, given by Eq.~(\ref{eq:hsstate}) above, has many symmetries and properties that we can use for constructing the optimal measurement.  Here we review these properties, and the reader is referred to older papers like ~\cite{Bacon:06f} where these properties are described in more detail.

First, define the left and right regular representations of ${\mathcal G}$.  The left regular representation is the representation of ${\mathcal G}$ given by $D_L(g^\prime)|g\rangle=|g^\prime g\rangle$.  Similarly the right regular representation is defined by $D_R(g^\prime)|g\rangle=|g (g^\prime)^{-1} \rangle$.  The hidden subgroup state can, because it is symmetric under the left regular representation, $D_L(g) \rho_{\mathcal H} D_L(g^{-1})=\rho_{\mathcal H}$, be written as a sum over right regular representation elements.  In fact it can be written in the particularly simple form
\begin{equation}
\rho_{\mathcal H}={1 \over |{\mathcal G}|} \sum_{h \in {\mathcal H}} D_R(h).
\end{equation}
If we define the subgroup projection operator
\begin{equation}
P_{\mathcal H}:={1 \over |{\mathcal H}|} \sum_{h \in {\mathcal H}} D_R(h),
\end{equation}
then we can express the hidden subgroup state as $\rho_{\mathcal H}=\frac{|{\mathcal H}|}{|{\mathcal G}|} P_{\mathcal H}$.  The reason for doing this is to notice that $P_{\mathcal H}$ is a projector, $P_{\mathcal H}^2=P_{\mathcal H}$ and to further note the role of this projector in the construction of the optimal measurements like those constructed by Ip~\cite{Ip:03a}.

While expressing the hidden subgroup state in terms of the right regular representation is helpful, it is even more helpful to examine this state in a different basis.  If we perform the quantum Fourier transform over the finite group ${\mathcal G}$, this transforms the basis of group elements $\{|g\rangle\}$ to a basis with irreducible representation (irrep) labels along with row and column registers, $\{|\mu,i,j\rangle\}$.  In this basis the hidden subgroup state is
\begin{equation}
\rho_{\mathcal H}=\bigoplus_\mu  \left[ I_{d_\mu} \otimes {1 \over |{\mathcal G}|} \sum_{h \in {\mathcal H}} D_{\mu}(h) \right],
\end{equation}
where $\mu$ labels the irreducible representations of ${\mathcal G}$, $D_\mu$ is the $\mu$th irrep, $d_\mu$ is the dimension of the $\mu$th irrep, and $I_{d_\mu}$ is the $d_\mu$ dimensional identity operator.  Defining the subgroup projectors $P_{\mathcal H}$ for the $\mu$th irrep,
\begin{equation}
P_{\mu,\mathcal H}:={1 \over |{\mathcal H}|} \sum_{h \in {\mathcal H}} D_\mu(h),
\end{equation}
we can express the hidden subgroup state as
\begin{equation}
\rho_{\mathcal H}=\bigoplus_\mu \left[ I_{d_\mu} \otimes \frac{|{\mathcal H}|}{|{\mathcal G}|} P_{\mu,\mathcal H} \right].
\end{equation}

A final useful symmetry relationship that we will use later comes about when we sum the projectors $P_{\mu,\mathcal H}$ over all conjugate subgroups.  A conjugate subgroup of ${\mathcal H}$ is defined as ${\mathcal H}^g:=\{g hg^{-1},h \in {\mathcal H}\}$ for $g \in {\mathcal G}$.  In particular we recall that
$ \sum_{g \in {\mathcal G}} D_{\mu} (g h g^{-1})$ commutes with all $D_{\mu}(g^\prime)$ and thus, via Schur's lemma, it must be proportional to the identity matrix (this proportionality may be zero),
\begin{equation}
 \sum_{g \in {\mathcal G}} D_{\mu} (g h g^{-1}) = r_\mu(h) I_{d_\mu},
\end{equation}
where $r_\mu(h)$ is the proportionality constant.  Taking the trace of this equation allows us to calculate $r_\mu(h)$:
\begin{equation}
\sum_{g \in {\mathcal G}} \chi_{\mu} (g h g^{-1}) =  |{\mathcal G}| \chi_{\mu} (h ) = r_\mu(h) d_\mu
\end{equation}
where $\chi_\mu(g):={\rm Tr}[D_\mu(g)]$ is the character of element $g$ in irrep $\mu$.  Thus we obtain
\begin{equation}
{1 \over |{\mathcal G}|} \sum_{g \in {\mathcal G}} P_{\mu,{\mathcal H}^g}= {1 \over |{\mathcal H}|} \sum_{h \in {\mathcal H}} { \chi_{\mu}(h) \over {d_\mu}} I_{d_\mu}. \label{eq:sump}
\end{equation}

\subsection{The Optimal Single Copy Measurement}

The optimal measurement criteria, Eq.~(\ref{eq:optimal}), when all of the subgroups are given with equal a prior probability is given by
\begin{eqnarray}
\sum_{{\mathcal H} \in {\rm Sub}({\mathcal G})} \rho_{{\mathcal H} } E_{\mathcal H} &=& \sum_{{\mathcal H} \in {\rm Sub}({\mathcal G})} E_{\mathcal H} \rho_{\mathcal H} \nonumber \\
\sum_{{\mathcal H} \in {\rm Sub}({\mathcal G})} \rho_{\mathcal H} E_{\mathcal H} &\geq& \rho_{{\mathcal H}^\prime}, \quad {\rm for~all~}{{\mathcal H}^\prime}\in {\rm Sub}({\mathcal G}),
\end{eqnarray}
where this sum over ${\mathcal H}$ is over all possible subgroups of ${\mathcal G}$.
Notice that if $\rho_{\mathcal H}$ and $E_{\mathcal H}$ commute, then the first equation is satisfied.  This will be true of the measurements we will consider, so the second equation will be the nontrivial condition.

Our ansatz for the optimal measurement is a measurement of the form
\begin{equation}
E_{\mathcal H}=\bigoplus_\mu \left[ I_{d_\mu} \otimes c_{\mu,{\mathcal H}}  P_{\mu,{\mathcal H}} \right].
\end{equation}
where $c_{\mu,{\mathcal H}}$ are as of yet undetermined constants.  The reason for beginning with this assumption is that it can easily be shown that the previous results of Ip~\cite{Ip:03a} and Moore and Russell~\cite{Moore:05c} can be written as measurements of this form.  Notice that this choice of measurement implies that $\rho_{\mathcal H}$ and $E_{\mathcal H}$ commute so the first optimal measurement criteria is automatically satisfied.  Thus we must show that the second measurement criteria is satisfied, while the condition that the measurement is valid also holds.  The $E_{\mathcal H}$ of our chosen form are a valid measurement if and only if $c_{\mu,{\mathcal H}} \geq 0$ and, for all $\mu$,
\begin{equation}
\sum_{{\mathcal H} \in {\rm Sub}({\mathcal G})} c_{\mu,{\mathcal H}} P_{\mu,{\mathcal H}}= I_{d_\mu}.
\end{equation}

It will be convenient to break the sum over all subgroups that appears here down into a sum over conjugate subgroups.  In particular two subgroups ${\mathcal H}_1$ and ${\mathcal H}_2$ are conjugate if there exists an element, $g$ of ${\mathcal G}$, such that ${\mathcal H}_2=\{ g h g^{-1}, h \in {\mathcal H}_1\}$.  Whether two subgroups are conjugate to each other forms an equivalence relation and thus we can partition the set of all subgroups of ${\mathcal G}$ into sets of conjugate subgroups.  Let ${\bf C}({\mathcal G})$ be the set of all of these sets of conjugate subgroups.  Then a sum over all subgroups ${\mathcal H}$ of ${\mathcal G}$ can be broken down into a sum over different sets of conjugate subgroups, i.e.
\begin{equation}
\sum_{{\mathcal H} \in {\rm Sub}({\mathcal G})}= \sum_{C \in {\bf C}({\mathcal G})} \sum_{{\mathcal H} \in C}.
\end{equation}

Applying this to the condition that the $E_{\mathcal H}$ are valid measurement operators yields,
\begin{equation}
\sum_{C \in {\bf C}({\mathcal G})} \sum_{{\mathcal H} \in C} c_{\mu,{\mathcal H}} P_{\mu,{\mathcal H}}= I_{d_\mu}
\end{equation}
or, explicitly performing the sum over conjugate subgroups,
\begin{equation}
\sum_{C \in {\bf C}({\mathcal G})}  \sum_{g \in {\mathcal G}} {|C| \over |{\mathcal G}|} c_{\mu,g{\mathcal H}_C g^{-1}} P_{\mu,g {\mathcal H}_C g^{-1}} = I_{d_\mu}
\end{equation}
where $|C|$ is the number of the subgroups in $C$, ${\mathcal H}_C$ is a representative subgroup of $C$ and
$P_{\mu,g {\mathcal H}_C g^{-1}}$ is the $\mu$th irrep projector onto the subgroup $g {\mathcal H}_C g^{-1}= \{g h g^{-1}, h \in {\mathcal H}_C\}$.

We now make the further assumption that $c_{\mu,{\mathcal H}}$ depends only on which $C \in {\bf C}({\mathcal G})$ the ${\mathcal H}$ belongs to.  Then we obtain
\begin{equation}
\sum_{C \in {\bf C}({\mathcal G})}{|C| \over |{\mathcal G}|} c_{\mu,C} \sum_{g \in {\mathcal G}} P_{\mu,g {\mathcal H}_C g^{-1}}= I_{d_\mu}
\end{equation}
Now exploit the fact that we have symmetrized the projectors over the group ${\mathcal G}$.  This is the trick relevant to the construction of Moore and Russell where they demonstrate that the PGM is optimal for subgroups selected uniformly from a set of conjugate subgroups~\cite{Moore:05c}.  While the case we consider includes a symmetrized sum like that used in ~\cite{Moore:05c}, our more general setting also has a sum over the different sets of conjugate subgroups.  Using Eq.~(\ref{eq:sump}) we find that
\begin{equation}
{1 \over d_\mu} \sum_{C \in {\bf C}({\mathcal G})} {|C| \over |{\mathcal H}_C|} c_{\mu,C}  \sum_{h \in {\mathcal H}_C } \chi_\mu(h)  = 1,
\end{equation}
where $|{\mathcal H}_C|$ is the number of elements in the subgroup ${\mathcal H}_C$.

Now back to the nontrivial inequality for the optimal measurement.  This is, for all ${\mathcal H}^\prime \in {\rm Sub}({\mathcal G}) $ and for all irrep labels $\mu$,
\begin{equation}
\sum_{{\mathcal H}\in{\rm Sub}({\mathcal H})}  |{\mathcal H}|c_{\mu,{\mathcal H}} P_{\mu,{\mathcal H}}  \geq |{\mathcal H}^\prime|P_{\mu,{\mathcal H}^\prime}.
\end{equation}
Turning the sum on the left hand side into a sum over conjugate subgroups this becomes
\begin{equation}
 \sum_{C \in {\bf C}({\mathcal G})} \sum_{{\mathcal H} \in C}
 |{\mathcal H}| c_{\mu,C} P_{\mu,{\mathcal H}}  \geq |{\mathcal H}^\prime|P_{\mu,{\mathcal H}^\prime}
\end{equation}
or
\begin{equation}
\sum_{C \in {\bf C}({\mathcal G})} \sum_{g \in {\mathcal G}}
 {|{\mathcal H}_C| |C| \over |{\mathcal G}|} c_{\mu,C} P_{\mu,g {\mathcal H}_C g^{-1}}  \geq |{\mathcal H}^\prime|P_{\mu,{\mathcal H}^\prime}
\end{equation}
Using our expression for the sum over the conjugate projectors, Eq.~(\ref{eq:sump}), we obtain
\begin{equation}
\sum_{C \in {\bf C}({\mathcal G})}
|C| {c_{\mu,C} \over d_\mu} \sum_{h \in {\mathcal H}_C} {\chi_\mu(h) } I_{d_\mu}  \geq |{\mathcal H}^\prime|P_{\mu,{\mathcal H}^\prime}
\end{equation}

Defining
\begin{equation}
s_\mu({\mathcal H}_C):={|C| \over d_\mu |{\mathcal H}_C|} \sum_{h \in {\mathcal H} } \chi_\mu (h),
\end{equation}
we can then write the condition that the generalized measurement is a valid measurement as
\begin{equation}
\sum_{ C \in {\bf C}({\mathcal G})} s_\mu({\mathcal H}_C) c_{\mu,C} =1, \quad c_{\mu,C} \geq 0 \label{eq:opt1}
\end{equation}
and the optimality condition as
\begin{equation}
\sum_{C \in {\bf C}({\mathcal G})}
|{\mathcal H}_C|  s_{\mu}({\mathcal H}_C ) c_{\mu,C}   I_{d_\mu}\geq |{\mathcal H}^\prime|P_{\mu,{\mathcal H}^\prime} \label{eq:opt2}
\end{equation}
for all ${\mathcal H}^\prime \in {\rm Sub}({\mathcal G})$.  Note that $s_{\mu}({\mathcal H})_C$ is a non-negative real number.  Equations (\ref{eq:opt1}) and (\ref{eq:opt2}) represent the optimality criteria for our ansatz.  We will now show that it is possible to choose a particular $c_{\mu,C}$ that satisfies both of these equations and thus we will have identified the optimal measurement.

Begin by examining the condition of Eq.~(\ref{eq:opt2}) for a fixed ${\mathcal H}^\prime$.  If $P_{\mu,{\mathcal H}^\prime}=0$, then this condition is automatically satisfied.  Thus we can assume $P_{\mu,{\mathcal H}^\prime} \neq 0$.  Now we will choose the constant $c_{\mu,C}$ that yields the optimal measurement.  Pick the largest subgroup ${\mathcal H}$ such that $P_{\mu,{\mathcal H}} \neq 0$ (break ties arbitrarily) and call it ${\mathcal H}_{\rm max}$.  This subgroup will belong to a particular set of conjugate subgroups, call this $C_{\rm max}(\mu)$.  Then set
\begin{equation}
c_{\mu,C}=\left\{\begin{array}{ll} {1 \over s_{\mu}({\mathcal H}_C)}  &{\rm if~}C=C_{\rm max}(\mu)\\  0 & {\rm if~}C \notin C_{\rm max}(\mu) \end{array} \right.
\end{equation}
(Note that $s_\mu({\mathcal H}_C)$ is not zero because of our condition that $P_{\mu,{\mathcal H}^\prime} \neq 0$.)  Certainly this expression for $c_{\mu,C}$ obeys the normalized measurement condition and is positive.  Further, since ${\mathcal H}_{\rm max}$ is the largest subgroup with $P_{\mu,{\mathcal H}} \neq 0$ all of the inequalities of Eq.~(\ref{eq:opt2}) will also hold.  This is because on the left hand side we have the identity operator and on the right hand side we have a projector and since these two are both diagonal in a fixed basis, the inequalities turns into an inequality between the constants appearing before these operators.  Since $|{\mathcal H}_{\rm max}|$ is larger than all $|{\mathcal H}|$ for which $P_{\mu,{\mathcal H}} \neq 0$, it follows that these constants in front of the operators obey the inequality.  Thus we have shown that this measurement is the optimal measurement for the single copy hidden subgroup problem.

It is useful to rephrase the optimal measurement we have derived.  The optimal measurement consists of measurement operators of the form
\begin{equation}
E_{\mathcal H}= \bigoplus_{\mu} I_{d_\mu} \otimes e_{\mu,{\mathcal H}} P_{\mu,{\mathcal H}}
\end{equation}
where
\begin{equation}
e_{\mu,{\mathcal H}}= \left\{\begin{array}{ll} {|{\mathcal H}_{\rm max}| \over |C_{\rm max}(\mu)|} {d_{\mu} \over \sum_{h \in {\mathcal H}_{\rm max}} \chi_\mu(h)}& {\rm if~} {\mathcal H} \in C_{\rm max} (\mu) \\
0 & {\rm if~} {\mathcal H} \notin C_{\rm max}(\mu)\end{array} \right. .
\end{equation}

It is easy to see that in our derivation of the optimal single copy measurement, had we restricted ourselves to only one set of conjugate subgroups, we would have obtained the PGM, while if we had restricted ourselves to only subgroups belonging to different sets of conjugate subgroups we would have obtained a measurement like that of Ip~\cite{Ip:03a}.  Thus the optimal measurement for a single copy of the HSP is a hybrid between the previous two known single copy optimal measurements for the HSP.

\subsection{Strengthening the Result}

It is also possible to strengthen our result beyond the case where all subgroups are given with equal probability.  In particular consider the case where the probability of a subgroup is uniform across different conjugate subgroups, but allowed to vary between the different conjugate subgroups.  In such a setting, the a prior probability of a subgroup $p_{\mathcal H}$ depends only on which set of conjugate subgroups ${\mathcal H}$ belongs to.  In other words, consider a probability $p_{\mathcal H}$ such that $p_{\mathcal H}=p_{{\mathcal H}^\prime}$ if ${\mathcal H}$ and ${\mathcal H}^\prime$ are conjugate to each other.  Keep $p_{\mathcal H}$ throughout the calculations performed above leads to a set of conditions similar to Eqs.~(\ref{eq:opt1}) and (\ref{eq:opt2}).  In particular the condition that the generalized measurement is a valid measurement remains unchanged,
\begin{equation}
\sum_{ C \in {\bf C}({\mathcal G})} s_\mu({\mathcal H}_C) c_{\mu,C} =1, \quad c_{\mu,C} \geq 0
\end{equation}
but the optimality condition becomes
\begin{equation}
\sum_{C \in {\bf C}({\mathcal G})}
p_{{\mathcal H}_C}|{\mathcal H}_C|  s_{\mu}({\mathcal H}_C ) c_{\mu,C}   I_{d_\mu}\geq  p_{{\mathcal H}^\prime} |{\mathcal H}^\prime|P_{\mu,{\mathcal H}^\prime}
\end{equation}
for all ${\mathcal H}^\prime \in {\rm Sub}({\mathcal G})$.  From this expression it is clear that instead of choosing the largest ${\mathcal H}$ such that $P_{\mu,{\mathcal H}} \neq 0$, the above condition can be satisfied by choosing a subgroup ${\mathcal H}_*$ with $P_{\mu,{\mathcal H}_*} \neq 0$ such that $p_{{\mathcal H}_*} |{\mathcal H}_*|$ is maximal (breaking ties arbitrarily.)  Let $C_*$ denote the set of conjugate subgroups to which such a ${\mathcal H}_*$ belongs.  Given this choice the optimal measurement will have the form
\begin{equation}
E_{\mathcal H}= \bigoplus_{\mu} I_{d_\mu} \otimes e_{\mu,{\mathcal H}} P_{\mu,{\mathcal H}}
\end{equation}
where
\begin{equation}
e_{\mu,{\mathcal H}}= \left\{\begin{array}{ll} {|{\mathcal H}_{*}| \over |C_{*}(\mu)|} {d_{\mu} \over \sum_{h \in {\mathcal H}_{*}} \chi_\mu(h)}& {\rm if~} {\mathcal H} \in C_{*} (\mu) \\
0 & {\rm if~} {\mathcal H} \notin C_{*}(\mu)\end{array} \right. .
\end{equation}
Thus we have solved the slightly more general problem of the optimal measurement when the a priori probability of a subgroup is required to be constant only across conjugate subgroups.  Optimal measurements for arbitrary a priori probability distributions are likely to be more difficult to obtain, since the symmetry arguments we have used do not hold in such a setting.

\section{Conclusion and Outlook}

We have determined the optimal measurement for a single copy of the HSP when all possible subgroups of a group are given with equal a priori probability.  We have also presented a slightly stronger version of this problem where the a priori probabilities are uniform over conjugate subgroups.  An important open problem is to determine the optimal measurement in the multi-copy version of this problem for the case where subgroups are given with uniform a priori probability.  This is especially true since it is known that a multi-copy measurement is necessary for any algorithm that hopes to efficiently solve the HSP for the symmetric group ~\cite{Hallgren:06a}.  Recent results for the Heisenberg HSP~\cite{Bacon:06f,Bacon:06e,Bacon:06d} give us the intuition that this measurement must have a block diagonal form related to the Clebsch-Gordan transform over the group ${\mathcal G}$.

\acknowledgements

We would like to acknowledge Andrew Childs for useful conversations and an anonymous referee for catching an error in an early version of this paper.  DB and TD are supported under ARO/NSA quantum algorithms grant number W911NSF-06-1-0379.  DB is also supported under NSF grant number 0523359 and NSF grant number 0621621.

\end{document}